\documentclass{article}

\usepackage{arxiv}
\usepackage[table]{xcolor}
\PassOptionsToPackage{hyphens}{url}
\usepackage{hyperref}
\usepackage{booktabs}
\usepackage{amsfonts}
\usepackage{nicefrac}
\usepackage{microtype}      
\usepackage{lipsum}
\usepackage{graphicx}
\usepackage[numbers]{natbib}
\usepackage{ulem}
\usepackage{pgf, tikz}
\usetikzlibrary{arrows, automata}
\usepackage{epstopdf}
\usepackage{pgfplots}
\usepackage{multirow}
\usepackage{cleveref}
\usepackage{xcolor}
\usepackage{soul,color}
\usepackage{helvet}
\usepackage{courier}
\usepackage{arabtex}
\usepackage{utf8}
\usepackage{adjustbox}

\title{Google Searches and COVID-19 Cases in Saudi Arabia: A Correlation Study}
\author{
  Btool Hamoui\thanks{Corresponding author} \\
 Center of Innovation and Development in Artificial Intelligence\\
 Umm Al-Qura University\\
 Makkah, Saudi Arabia \\
  \texttt{s43680523@st.uqu.edu.sa} \\
  \AND
 Abdulaziz Alashaikh \\
  Computer and Networks Engineering Department\\
  University of Jeddah\\
  Jeddah, Saudi Arabia \\
  \texttt{asalashaikh@uj.edu.sa} \\
  \AND
  Eisa Alanazi \\
  Center of Innovation and Development in Artificial Intelligence \\
  Umm Al-Qura University\\
  Makkah, Saudi Arabia \\
  \texttt{eaanazi@uqu.edu.sa} \\
}

\let\oldmaketitle\maketitle
\renewcommand{\maketitle}{\oldmaketitle\setcounter{footnote}{0}}

\begin{document}
\maketitle
\begin{abstract}
\textbf{Background}:
The outbreak of the new coronavirus disease (COVID-19) has affected human life to a great extent on a worldwide scale. During the coronavirus pandemic, public health professionals at the early outbreak faced an extraordinary challenge to track and quantify the spread of disease.

\textbf{Objective}:
To investigate whether a digital surveillance model using google trends (GT) is feasible to monitor the outbreak of coronavirus in the Kingdom of Saudi Arabia.

\textbf{Methods}:
We retrieve GT data using ten common COVID-19 symptoms related keywords from March 2, 2020, to October 31, 2020. Spearman correlation were performed to determine the correlation between COVID-19 cases and the Google search terms.

\textbf{Results}: GT data related to Cough and Sore Throat were the most searched symptoms by the Internet users in Saudi Arabia. The highest daily correlation found with the Loss of Smell followed by Loss of Taste and Diarrhea.
Strong correlation as well was found between the weekly confirmed cases and the same symptoms: Loss of Smell, Loss of Taste and Diarrhea.

\textbf{Conclusions}:
 We conducted an investigation study utilizing Internet searches related to COVID-19 symptoms for surveillance of the pandemic spread. This study documents that google searches can be used as a supplementary surveillance {tool} in COVID-19 monitoring in Saudi Arabia.

 \end{quote}
\end{abstract}


\section{Introduction}
Coronavirus was first reported in December 2019 in China, then {has continuously and gradually} spread in several different countries and became a global pandemic in March 2020. In response to the COVID-19 pandemic, countries have different governance mechanism applied to combat the pandemic. The pandemic has affected human life and triggered alerts around the globe. However, the responses toward it were mainly depend on the local governance. In Saudi Arabia, strict control and preventive measures were undertaken to prevent the spread of the outbreak. Although, the spread of the virus across the kingdom is considered limited comparing to other countries  \cite{algaissi2020preparedness}; there is a critical need to activate digital health surveillance to make the response more effective and reduce the risk of further spread of the disease.

In the past few years, the Internet has become a very popular medium for people searching for health-related knowledge and information for self-diagnosis. During the coronavirus emerging, people opt to search about the disease signals, such as the appearance of specific symptoms, the treatments, and the recovery of those symptoms on the Internet. Consequently, Internet searches are considered important user-generated content which may include information about user health-related information. The social media posts content and Internet search data can play a prominent role to promote health situations during an emerging outbreak \cite{zhang2019social}. The systems or applications that use Internet-based data for the purpose of nowcasting or forecasting disease infections known as digital disease surveillance \cite{aiello2020social}. 

Numerous works have proposed disease surveillance methodologies by taking the advantage of user-generated web content, either in the form of social media posts or search engine query logs. In 2008, Google developed an early warning system that predicts influenza activity by aggregating Google search query volumes related to flu symptoms \cite{ginsberg2009detecting,wilson2009early}. During Zika outbreak in 2016 \cite{mcgough2017forecasting}, a predictive model developed utilizing tweets posts, google trends, and HealthMap reports. The model achieved a successful prediction of Zika cases counts in Latin America compared to traditional surveillance systems.

Recently, several research attempts have exploited the information in google trends  to better monitor the recent COVID-19 outbreak \cite{walker2020use,venkatesh2020prediction,lin2020google}. 
The study by \citet{walker2020use} demonstrated a strong correlation between the google searches related to smell information and COVID-19 cases in Italy, Spain, UK, USA, Germany, France, Iran, and the Netherlands.
The correlation analysis between google search keywords “coronavirus”, “COVID”, “COVID 19”, “corona”, and “virus” with daily confirmed cases in india presented in   \cite{venkatesh2020prediction}. Another study examined google searches keywords “wash hands” and “face mask” correlation with number of confirmed cases among 21 countries \cite{lin2020google}.   

Saudi Arabia has the largest Internet user population in the Arab world  \cite{simsim2011internet}. An investigation done by  \citet{alduraywish2020sources} found that the Internet is one of the most common sources to retrieve health information { for Saudis}. The study aimed to validate utilizing Google Trends data as a complement data source for digital surveillance in Saudi Arabia.
{ In this work, we investigate the correlation between Google Trends data on coronavirus common symptoms using Arabic keywords and the COVID-19 confirmed cases in Saudi Arabia.}

%

\section{Methods}
\label{sec:Methods}

\begin{figure}[htp]
    \centering
    \includegraphics[width=.9\textwidth ]{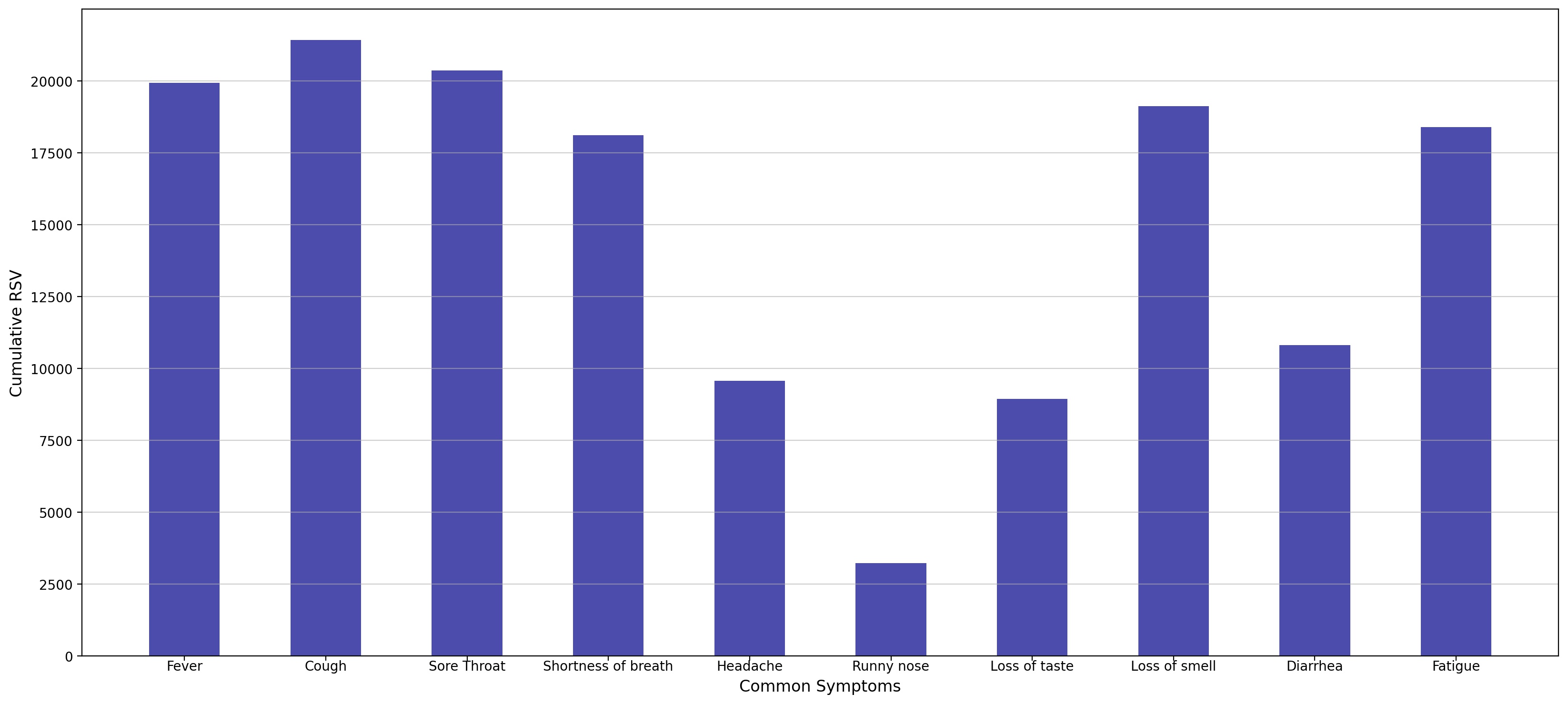}
    \caption{Relative search volumes (RSVs) of trending  Google symptoms keywords }
    \label{fig:RSV}
\end{figure}

\setcode{utf8}
The study period was from March 2, 2020 to October 31, 2020. March 2, 2020 is the symptom onset day of the first confirmed case with a positive coronavirus result. Data of daily COVID-19 cases in Saudi Arabia were collected from the Ministry of Health (MOH) daily reports\footnote{available at: \url{https://datasource.kapsarc.org/explore/dataset/saudi-arabia-coronavirus-disease-covid-19-situation/table/?disjunctive.daily_cumulative&disjunctive.indicator&disjunctive.event&disjunctive.city_en&disjunctive.region_en&sort=date}}. 
We manually crafted a list of 26 Arabic {n-grams} keywords related to coronavirus symptoms as shown in \Cref{table:Symptomskeyterms}. We obtained the symptoms from Google Search trends, using Google Trends, an open-access platform that provides the relative search volume (RSV)\footnote{available at: \url{https://trends.google.com/trends/?geo=SA}} (scaled search frequency data from 0 to 100). The daily trend data associated with the list of Arabic keywords obtained from Google Trends by setting the location parameter to “Saudi Arabia” and the time parameter to “March to October, 2020.”. The total number of RSVs for each symptom is shown in \Cref{fig:RSV}. The strengths of the associations between both daily and weekly increase of confirmed cases, and google trends search queries will be assessed using the Spearman rank correlation. An r-value of $>$ 0.5 is considered as a high correlation, and a p-value of $<$ 0.05 is considered as a statistically significant result. We tested the correlation of each 29 symptoms keywords and the overall number for each RSV symptom. The results of daily and weekly correlations of overall RSV for each symptoms with COVID-19 cases are presented in \Cref{table:GooglesearchesAr}. 

\begin{table*}[h]
\centering
\resizebox{\textwidth}{!}{%
\begin{tabular}{ c|c|c}
 \hline
 Symptoms in English & Related Symptoms list  & Symptoms in Arabic  \\
 \hline\hline
Fever & 
 \<حمى ,ارتفاع درجة حرارة الجسم , ارتفاع حرارة الجسم >
 & \<الحمى>\\ 
 Cough &  
 \<سعال ، كحة ، السعال الجاف ، الكحة الجافة>

 & \<سعال>\\
 Fatigue &
\< إعياء, تعب, إرهاق>
 &\<إعياء>\\
Sore Throat & 
\< التهاب الحلق ، التهاب في الحلق ، الام الحلق>
 & \<التهاب الحلق >\\
  shortness of breath & \< ضيق التنفس, ضيق في التنفس ,>
& \<ضيق التنفس>\\
  Headache & 

\<صداع >
 & \<صداع>\\
Anosmia &  
\<فقدان حاسة الشم ,فقدان الشم , فقد حاسة الشم ,فقد الشم ,  >
 & \<فقدان حاسة الشم>\\
 Loss of taste &
\< فقدان حاسة التذوق, فقدان التذوق,فقد حاسة التذوق, فقد التذوق, >
 &\<فقدان حاسة التذوق>\\
Diarrhea & 
\<إسهال  >
 & \<إسهال>\\
  Runny nose & 
   \< سيلان الأنف >

 & \<سيلان الأنف>\\
\hline
\end{tabular}
}
\caption{List of symptoms key terms }
\label{table:Symptomskeyterms}
\end{table*}

\section{Results and Discussion}
\label{sec:Results}


\begin{figure}[htp]
    \centering
    \includegraphics[width=.85\textwidth , height=.9\textheight]{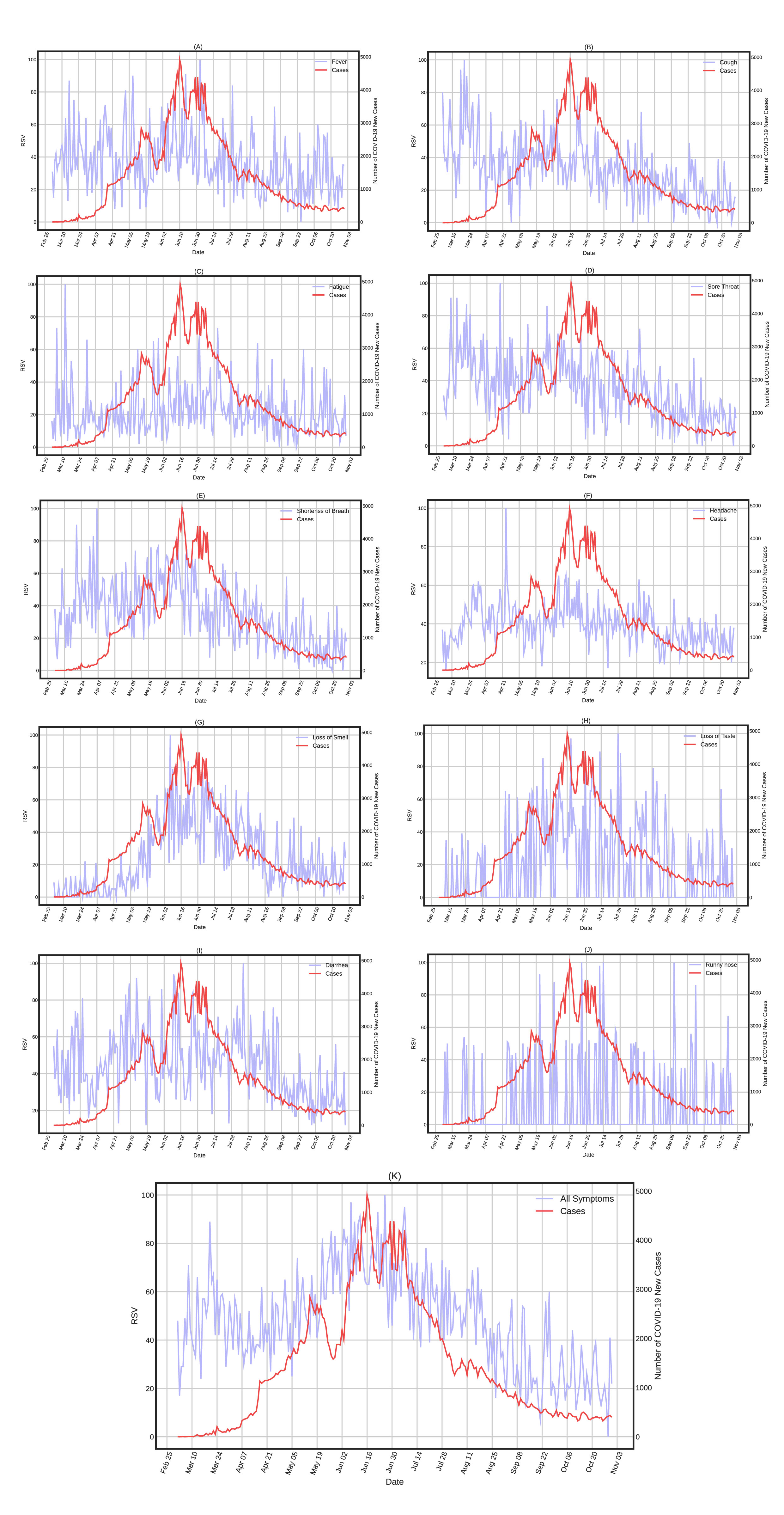}
    \caption{Relative Search Volume (RSV) for Fever (A), Cough (B), Fatigue (C), Sore throat (D), shortness of breath (E), Headache (F), Loss of Smell (G), Loss of Taste (H), Diarrhea (I), Runny Nose (J) and the total GT symptoms data (K) with confirmed COVID-19 cases.}
    \label{fig:RSVCases}
\end{figure}

During the period between {March 2 and October 31}, it is observed from the \Cref{fig:RSV} that the highest google searches queries were about Cough, Sore Throat, and Fever. The RSVs for each of them reach 21,428, 20,376 and 19,934, respectively. Regarding daily Spearman correlation, the Arabic searches about ``Loss of Smell'' were strongly correlated and statistically significant with COVID-19 cases as shows in the \Cref{table:GooglesearchesAr}. Moreover, moderate correlations were observed with symptoms related to ``Loss of Taste'', ``Diarrhea'', and ``shortness of breath'' with p-value $<$ 0.05. Although weak correlations were found with RSVs about ``Fever'', ``Headache'', ``Sore Throat'', and ``Fatigue'', the associations are statistically significant with p-value $<$0.05. In terms of weekly Spearman correlation, the RSVs pertaining ``Loss of Smell'', ``Loss of Taste'' and ``Diarrhea'' have a strong correlation ranging from 0.578 to 0.83. All three correlations were statistically significant ( p-value $<$ 0.05). Besides, moderate correlations with statistically significant ( p-value $<$ 0.05) found with ``Shortness of Breath'', ``Headache'', ``Fatigue'' and ``Fever''. 
The correlations of overall symptoms were strongly correlated and statistically significant with both daily and weekly COVID-19 cases. 
\Cref{fig:RSVCases} shows the daily RSVs of ``Loss of Smell'', ``Loss of Taste'', and the overall symptoms with the confirmed COVID-19 cases. We observed that the ``Loss of Smell'' searches increase and decrease simultaneity with the daily COVID-19 cases. 
%
%
\begin{table*}[!t]
\caption{Spearman Correlation Coefficients Relating Arabic Google searches with COVID-19 incidence in Saudi Arabia.}
\label{table:GooglesearchesAr}
\centering
\begin{tabular}{ccccc}
\hline

\multicolumn{1}{c}{\multirow{2}{*}{Google Searches about}} & \multicolumn{2}{c}{Daily Correlation} & \multicolumn{2}{c}{Weekly Correlation} \\  
\multicolumn{1}{c}{}                           & r-value            & p-value              & r-value             & p-value              \\  \\ \hline\hline \cline{1-1} \hline\hline 
Cough                      & 0.118         & 0.065                & 0.135         & 0.436                \\  \\ \hline\hline
Fever                      & 0.167         & 0.008                & 0.366           & 0.030                \\  \\ \hline\hline
Sore Throat                & 0.170         & 0.007                & 0.169          & 0.329                \\  \\ \hline\hline
Shortness of Breath        & 0.343          & $1.27\times 10^{-3} $              & 0.438           & 0.008                \\  \\ \hline\hline
Headache                   & 0.298          & $5.13\times 10^{-3}$               & 0.416           & 0.012                \\  \\ \hline\hline
Runny Nose                 & 0.109          & 0.088               & 0.331           & 0.051                \\  \\ \hline\hline
Fatigue                    & 0.240          & 0.00014               & 0.379           & 0.024                \\  \\ \hline\hline
Diarrhea                   & 0.378          & $1.58\times10^{-4}$       & \bf{0.578}           & 0.0002                \\  \\ \hline\hline
Loss of Smell              & \bf{0.705}          & $1.35 \times10^{-16}$        & \bf{0.830}           & $3.12 \times 10^{-4} $         \\  \\ \hline\hline
Loss of Taste              & 0.405          & $7.59\times10^{-5}$          & \bf{0.773}           & $1.68\times 10^{-3}$          \\  \\ \hline\hline
All Symptoms                                     & \bf{0.588}          & $1.42 \times 10^{-10}$         & \bf{0.725}           & $7.39\times 10^{-3}$    
\\\hline
\end{tabular}
\end{table*}


In Saudi Arabia, the number of daily confirmed cases reached its peak (4,919 cases) on June $16^{th}$ 2020, then it started decreasing for six days until June $23^{rd}$. From this date, the number of daily confirmed cases began to increase again to peak on June $29^{th}$, see \Cref{fig:RSVCases}. A decrease in the number of confirmed cases started on July $7^{th}$ and showed a continuous reduction for the rest of three months, August, September, and November. Consequently, the number of total confirmed cases reached its highest in June (107,083 cases), while the total number for July, August, September, and November were 87,783, 40,602, 19,366, and 12,693, respectively. 
Similarly, over the eight months, the highest total RSVs for (GT) related to ``Loss of Smell'', ``Shortness of Breath'', ``Fever'', and ``All symptoms'' the overall symptoms were found in June 2020 as shown in\Cref{fig:RSVCases}(G),(A) and (E) .

In late March 2020, the international medical community began circulating press releases of the loss of sense of smell as a sign of COVID-19, and possible markers of infection \cite{losssmell}. By the end of April 2020, the Centers for Disease Control and Prevention (CDC) added the ``Loss of Smell'' to the list of common symptoms of coronavirus \cite{center}. In addition, multiple studies {have identified} ``Loss of Smell'' or ``Anosmia'' as a prominent symptom of COVID-19 infection \cite{marchese2020loss,meng2020covid,alanazi2020jmir}. 
Our findings are consistent with recent studies that demonstrated the association of google searches related to ``Loss of Smell'' and COVID-19 cases \cite{walker2020use,cherry2020loss,kurian2020correlations}. Between February and May 2020, strong correlations found (r$>$0.65 ) between google searches of the term ``Loss of sense of smell'' and COVID-19 cases in Brazil, Italy, USA, France, and Spain \cite{cherry2020loss}. The term ``Loss of Smell'' was one of the ten keywords used to investigate the association between google searches and COVID-19 cases in United States \cite{kurian2020correlations}. In the period between January 22 and April 6 2020, the correlation was (r = 0.61) for the whole United States, while strong correlations found in New York and Arizona equal to (r =0.70) with lag -8 and (r= 0.66) with lag -3.  

Overall, the correlations found in our study demonstrate the utility of utilizing google searches in providing helpful data to be used in syndromic surveillance during an emerging pandemic such as COVID-19. The analysis showed the importance of taking advantages of google searches data. This type of data is easily accessible and available for free; it would augment and complement traditional public health disease surveillance. Since the data are made available to the public in real time, this will help health authorities to take the right action at the right time. 
However, there are limitations in the study presented here. For Internet-based data, the changes over time might be in response to media change during the outbreak. This is known as the media-driven bias that impacts internet-based surveillance systems, as reported by a previous study \cite{althouse2011prediction}. 
At the early stage of pandemic, the media and medical community have paid significant attention to certain symptoms such as Fever, Cough and Sore Throat. Hence, we justify the result of weak correlation with symptoms such as "cough" and "sore throat", that the awareness of individuals changed in accordance with media and news about the COVID-19 pandemic. As illustrated in \Cref{fig:RSVCases}(B) and \Cref{fig:RSVCases}(D), the most google searches about cough and sore throat were found in March. Additionally, one of the limitations of the study is coverage, google searches are not considered to be representative of the entire Saudi Arabia population. We only focus on the Arabic language as it is the most spoken language in Saudi Arabia, where there are other languages spoken by the residents such as English, Urdu and Indonesian. Furthermore, the residents of villages and remote areas that suffer from slow Internet connections or limited access do not have the opportunity to surf the Internet as others in urban area. 

Despite the limitations, as presented in previous studies \cite{cervellin2017google,barros2020application,higgins2020correlations}, we observed in our study that the symptoms of COVID-19 with less media coverage, which are: Loss of Smell, Loss of Taste and Diarrhea, can provide indications of virus spread. Hence, we still found that Internet searches are a viable data source and can be utilized as digitalized surveillance technique to assist public health efforts.

\section{Conclusions}
\label{sec:conclusions}
In this paper, we investigated the feasibility of using Google searches to track COVID-19 outbreak in Saudi Arabia. Our study showed the the overall Google Trends data about symptoms were strongly correlated with COVID-19 confirmed cases, and highly correlated with ``Loss of Smell" in particular. Our study demonstrated the potential role of using the Internet searches in the fight against the current pandemic. The study also highlights the advantages and limitations of Internet based data for digital health surveillance in Saudi Arabia.

\bibliographystyle{IEEEtranN}
\bibliography{reference} 

\begin{thebibliography}{22}
\providecommand{\natexlab}[1]{#1}
\providecommand{\url}[1]{#1}
\csname url@samestyle\endcsname
\providecommand{\newblock}{\relax}
\providecommand{\bibinfo}[2]{#2}
\providecommand{\BIBentrySTDinterwordspacing}{\spaceskip=0pt\relax}
\providecommand{\BIBentryALTinterwordstretchfactor}{4}
\providecommand{\BIBentryALTinterwordspacing}{\spaceskip=\fontdimen2\font plus
\BIBentryALTinterwordstretchfactor\fontdimen3\font minus
  \fontdimen4\font\relax}
\providecommand{\BIBforeignlanguage}[2]{{%
\expandafter\ifx\csname l@#1\endcsname\relax
\typeout{** WARNING: IEEEtranN.bst: No hyphenation pattern has been}%
\typeout{** loaded for the language `#1'. Using the pattern for}%
\typeout{** the default language instead.}%
\else
\language=\csname l@#1\endcsname
\fi
#2}}
\providecommand{\BIBdecl}{\relax}
\BIBdecl

\bibitem[Algaissi et~al.(2020)Algaissi, Alharbi, Hassanain, and
  Hashem]{algaissi2020preparedness}
A.~Algaissi, N.~Alharbi, M.~Hassanain, and A.~Hashem, ``Preparedness and
  response to covid-19 in saudi arabia: Lessons learned from mers-cov,'' 2020.

\bibitem[Zhang and Centola(2019)]{zhang2019social}
J.~Zhang and D.~Centola, ``Social networks and health: new developments in
  diffusion, online and offline,'' \emph{Annual Review of Sociology}, vol.~45,
  pp. 91--109, 2019.

\bibitem[Aiello et~al.(2020)Aiello, Renson, and Zivich]{aiello2020social}
A.~E. Aiello, A.~Renson, and P.~N. Zivich, ``Social media--and internet-based
  disease surveillance for public health,'' \emph{Annual Review of Public
  Health}, vol.~41, pp. 101--118, 2020.

\bibitem[Ginsberg et~al.(2009)Ginsberg, Mohebbi, Patel, Brammer, Smolinski, and
  Brilliant]{ginsberg2009detecting}
J.~Ginsberg, M.~H. Mohebbi, R.~S. Patel, L.~Brammer, M.~S. Smolinski, and
  L.~Brilliant, ``Detecting influenza epidemics using search engine query
  data,'' \emph{Nature}, vol. 457, no. 7232, pp. 1012--1014, 2009.

\bibitem[Wilson and Brownstein(2009)]{wilson2009early}
K.~Wilson and J.~S. Brownstein, ``Early detection of disease outbreaks using
  the internet,'' \emph{Cmaj}, vol. 180, no.~8, pp. 829--831, 2009.

\bibitem[McGough et~al.(2017)McGough, Brownstein, Hawkins, and
  Santillana]{mcgough2017forecasting}
S.~F. McGough, J.~S. Brownstein, J.~B. Hawkins, and M.~Santillana,
  ``Forecasting zika incidence in the 2016 latin america outbreak combining
  traditional disease surveillance with search, social media, and news report
  data,'' \emph{PLoS neglected tropical diseases}, vol.~11, no.~1, p. e0005295,
  2017.

\bibitem[Walker et~al.(2020)Walker, Hopkins, and Surda]{walker2020use}
A.~Walker, C.~Hopkins, and P.~Surda, ``The use of google trends to investigate
  the loss of smell related searches during covid-19 outbreak,'' in
  \emph{International Forum of Allergy \& Rhinology}.\hskip 1em plus 0.5em
  minus 0.4em\relax Wiley Online Library, 2020.

\bibitem[Venkatesh and Gandhi(2020)]{venkatesh2020prediction}
U.~Venkatesh and P.~A. Gandhi, ``Prediction of covid-19 outbreaks using google
  trends in india: A retrospective analysis,'' \emph{Healthcare informatics
  research}, vol.~26, no.~3, pp. 175--184, 2020.

\bibitem[Lin et~al.(2020)Lin, Liu, and Chiu]{lin2020google}
Y.-H. Lin, C.-H. Liu, and Y.-C. Chiu, ``Google searches for the keywords of
  “wash hands” predict the speed of national spread of covid-19 outbreak
  among 21 countries,'' \emph{Brain, Behavior, and Immunity}, 2020.

\bibitem[Simsim(2011)]{simsim2011internet}
M.~T. Simsim, ``Internet usage and user preferences in saudi arabia,''
  \emph{Journal of King Saud University-Engineering Sciences}, vol.~23, no.~2,
  pp. 101--107, 2011.

\bibitem[Alduraywish et~al.(2020)Alduraywish, Altamimi, Aldhuwayhi, AlZamil,
  Alzeghayer, Alsaleh, Aldakheel, and Tharkar]{alduraywish2020sources}
S.~A. Alduraywish, L.~A. Altamimi, R.~A. Aldhuwayhi, L.~R. AlZamil, L.~Y.
  Alzeghayer, F.~S. Alsaleh, F.~M. Aldakheel, and S.~Tharkar, ``Sources of
  health information and their impacts on medical knowledge perception among
  the saudi arabian population: Cross-sectional study,'' \emph{Journal of
  Medical Internet Research}, vol.~22, no.~3, p. e14414, 2020.

\bibitem[UK(2020 March 20)]{losssmell}
\BIBentryALTinterwordspacing
E.~UK, \emph{Loss of sense of smell as marker of COVID-19 infection}, 2020
  March 20. [Online]. Available:
  \url{https://www.entuk.org/sites/default/files/files/Loss%20of%20sense%20of%20smell%20as%20marker%20of%20COVID.pdf}
\BIBentrySTDinterwordspacing

\bibitem[español(2020 March 21)]{center}
\BIBentryALTinterwordspacing
E.~español, \emph{Coronavirus (COVID-19) Information for Employees and
  Patients}, 2020 March 21. [Online]. Available:
  \url{https://www.vumc.org/coronavirus/latest-news/five-things-know-about-smell-and-taste-loss-covid-19}
\BIBentrySTDinterwordspacing

\bibitem[Marchese-Ragona et~al.(2020)Marchese-Ragona, Restivo, De~Corso,
  Vianello, Nicolai, and Ottaviano]{marchese2020loss}
R.~Marchese-Ragona, D.~A. Restivo, E.~De~Corso, A.~Vianello, P.~Nicolai, and
  G.~Ottaviano, ``Loss of smell in covid-19 patients: a critical review with
  emphasis on the use of olfactory tests,'' \emph{Acta Otorhinolaryngologica
  Italica}, vol.~40, no.~4, p. 241, 2020.

\bibitem[Meng et~al.(2020)Meng, Deng, Dai, and Meng]{meng2020covid}
X.~Meng, Y.~Deng, Z.~Dai, and Z.~Meng, ``Covid-19 and anosmia: A review based
  on up-to-date knowledge,'' \emph{American Journal of Otolaryngology}, p.
  102581, 2020.

\bibitem[Alanazi et~al.(2020)Alanazi, Alashaikh, Alqurashi, and
  Alanazi]{alanazi2020jmir}
E.~Alanazi, A.~Alashaikh, S.~Alqurashi, and A.~Alanazi, ``Identifying and
  ranking common covid-19 symptoms from tweets in arabic: Content analysis,''
  \emph{J Med Internet Res}, vol.~22, no.~11, p. e21329, Nov 2020.

\bibitem[Cherry et~al.(2020)Cherry, Rocke, Chu, Liu, Lechner, Lund, and
  Kumar]{cherry2020loss}
G.~Cherry, J.~Rocke, M.~Chu, J.~Liu, M.~Lechner, V.~J. Lund, and B.~N. Kumar,
  ``Loss of smell and taste: a new marker of covid-19? tracking reduced sense
  of smell during the coronavirus pandemic using search trends,'' \emph{Expert
  Review of Anti-infective Therapy}, pp. 1--6, 2020.

\bibitem[Kurian et~al.(2020)Kurian, Alvi, Ting, Storlie, Wilson, Shah, Liu,
  Bydon, et~al.]{kurian2020correlations}
S.~J. Kurian, M.~A. Alvi, H.~H. Ting, C.~Storlie, P.~M. Wilson, N.~D. Shah,
  H.~Liu, M.~Bydon \emph{et~al.}, ``Correlations between covid-19 cases and
  google trends data in the united states: A state-by-state analysis,'' in
  \emph{Mayo Clinic Proceedings}.\hskip 1em plus 0.5em minus 0.4em\relax
  Elsevier, 2020.

\bibitem[Althouse et~al.(2011)Althouse, Ng, and
  Cummings]{althouse2011prediction}
B.~M. Althouse, Y.~Y. Ng, and D.~A. Cummings, ``Prediction of dengue incidence
  using search query surveillance,'' \emph{PLoS Negl Trop Dis}, vol.~5, no.~8,
  p. e1258, 2011.

\bibitem[Cervellin et~al.(2017)Cervellin, Comelli, and
  Lippi]{cervellin2017google}
G.~Cervellin, I.~Comelli, and G.~Lippi, ``Is google trends a reliable tool for
  digital epidemiology? insights from different clinical settings,''
  \emph{Journal of epidemiology and global health}, vol.~7, no.~3, pp.
  185--189, 2017.

\bibitem[Barros et~al.(2020)Barros, Duggan, and
  Rebholz-Schuhmann]{barros2020application}
J.~M. Barros, J.~Duggan, and D.~Rebholz-Schuhmann, ``The application of
  internet-based sources for public health surveillance (infoveillance):
  systematic review,'' \emph{Journal of Medical Internet Research}, vol.~22,
  no.~3, p. e13680, 2020.

\bibitem[Higgins et~al.(2020)Higgins, Wu, Sharma, Illing, Rubel, Ting, and
  Alliance]{higgins2020correlations}
T.~S. Higgins, A.~W. Wu, D.~Sharma, E.~A. Illing, K.~Rubel, J.~Y. Ting, and
  S.~F. Alliance, ``Correlations of online search engine trends with
  coronavirus disease (covid-19) incidence: Infodemiology study,'' \emph{JMIR
  public health and surveillance}, vol.~6, no.~2, p. e19702, 2020.

\end{thebibliography}


\end{document}